\documentstyle{elsart}

\def\beq{\begin{equation}}
\def\eeq{\end{equation}}

\def\bea{\begin{eqnarray}}
\def\eea{\end{eqnarray}}
\def\ba{\begin{array}}                  
\def\ea{\end{array}}

\begin{document}
\begin{frontmatter}
\title{Quantum  Horizons
and Space-Time Non-Commutativity }
\thanks[talk]{Expanded version of a poster presented at the 5 European
Advanced Study Conference in Ancient Olympia, Greece, May 2004.}
\author{M. Martinis and
V. Mikuta-Martinis}
\address{Rudjer Bo\v skovi\' c Institute, Zagreb, Croatia}

\begin{abstract}
We study  dynamics of a scalar field in the near-horizon region
described by a static Klein-Gordon operator which is the
Hamiltonian of the system. The explicite construction of a time
operator near-horizon is given and its self-adjointness discussed.
\end{abstract}
\end{frontmatter}
\newpage
\section{Introduction}
Einstein's general theory of relativity [Nicolson,1981] describes
gravity as a curvature of spacetime caused by the presence of
matter. A black hole is formed if  a massive object (e.g. a star)
collapses into an infinitely dense state known as a singularity.
In this case, the curvature of spacetime becomes extreme and
prevents any light or other electromagnetic radiation from
escaping to infinity. The event horizon defines the boundary of a
black hole  behind which nothing, not even light, can escape.

Applying the Einstein Field Equations to collapsing stars, German
astrophysicist Kurt Schwarzschild introduced the critical radius
$R_S = 2GMc^{-2}$ for a given mass at which matter would collapse
into a singularity. The simplest three-dimensional geometry for a
black hole is a sphere (known as a Schwarzschild black hole), its
surface defines the event horizon.

 Therefore, in the standard General Relativity without Quantum
Mechanics (QM) a fundamental instability against collapse implies
the existence of black holes as stable solutions of Einstein's
equations. The QM and the Quantum Field Theory (QFT)  in the
curved spacetime with classical event horizon are, however,
troubled by the singularity at the horizon ['t Hooft, 2004]. This
problem may be solved by treating the black hole  as a quantum
state which implies that the energy of the black hole and its
corresponding time do not commute at the horizon [Bai and Yan,
2004]. In this picture we study the complex dynamics of a scalar
field in the near-horizon region described by a static
Klein-Gordon operator which in this case becomes  the Hamiltonian
of the system. We present the  explicite construction of the time
operator near-horizon  and discuss its self-adjointness [Martinis
and Mikuta,2003].

\section{Scalar Field in the Near-Horizon Region}

The Schwarzschild geometry of a static spherical black hole is
described by the metric

\begin{equation}
 ds^2 = - f(r)dt^2 + [f(r)]^{-1}dr^2 + r^2d\Omega ^2 ,
\end{equation}

where $rf(r) = r - 2M$ and  $d\Omega ^2 = d\theta ^2 + sin^2
\theta d\varphi ^2$. This metric  is minimally coupled to a scalar
field $\Phi (r,t,\Omega)$ through the  action ($c=1, \hbar=1$ and
$G = 1$):
\begin{equation}
S = - \frac{1}{2}\int d^4x \sqrt{- g}[g^{\mu \nu }\partial _{\mu
}\Phi \partial _{\nu}\Phi + m^2\Phi ^2],
\end{equation}
where $ g = det[g^{\mu \nu}]$. The equation of motion in the
black-hole gravitational background is
\begin{eqnarray}
(\Box_g - m^2)\Phi & = & \frac{1}{\sqrt{- g}}\partial _{\mu
}(\sqrt{- g}g^{\mu \nu }\partial _{\nu}\Phi ) - m^2\Phi \nonumber
\\ & = &  - \frac{1}{f} \Phi _{tt} + f \Phi _{rr} + (f ' + 2f/r)\Phi _{r} + \triangle _{\Omega}\Phi/r^2 - m^2\Phi
\\ & = & 0.\nonumber
\end{eqnarray}

 By separating the time and angular variables
\begin{equation}
\Phi (t,r,\Omega ) = e^{-it\omega}f^{-1/2}r^{-1}\psi
(r)Y_{lm}(\Omega ),
\end{equation}
the behavior of the Schroedinger-like equation near the horizon
can be studied by means of an expansion in the near-horizon
variable $ x = r - 2M $. In this variable, for small $x$, we
obtain  a conformally invariant   Schroedinger equation
\begin{equation}
[\frac{d^2}{dx^2} + (\frac{1}{4} + \Theta ^2)x^{-2}] \psi(x) = 0
\end{equation}

where  $\Theta  = 2M\omega $. This reduction to an effective
Schroedinger-like equation can be  described by an effective
attractive  potential $ V(x) = -(1/4  + \Theta ^2) x^{-2}$ which
is conformally invariant with respect to the near-horizon variable
$x$. The corresponding quantum Hamiltonian is
\begin{equation}
H =  (p^2 + \frac{g}{x^2} ),
\end{equation}
 where $ g = -2 (1/4  + \Theta ^2)$ is supercritical for all nonzero frequencies $\omega $.

\section{Time Operator and Spacetime Noncommutativity}

The singularity associated with horizon may be overcome by
assuming spacetime non-commutativity on the horizon, the so called
Quantum Horizon ['t Hooft, 2004], [ Bai and  Yan, 2004]. If a
black hole is considered as a quantum state, its energy $\hat{H}$
and its conjugate time $\hat{t}$ are expected to become conjugate
operators obeying
\begin{equation}
[\hat{H},\hat{t}] = i.
\end{equation}
Since $\hat{H}$ is x-coordinate dependent,  we expect spacetime
noncommutativity, $[\hat{t},x] \neq 0$. What is $\hat{t}$ as an
operator? Due to Pauli theorem [Pauli,1958] no such self-adjoint
operator should exist if the spectrum of the Hamiltonian is
semibounded or discrete. In quantum theory $\hat{H}$ has a
continuous spectrum for $g \ge 3/4$ with $E > 0$ but no ground
state at $E = 0$. It is  essentially self- adjoint only for $g >
3/4$ in  the domain
\begin{equation}
{\cal D}_0 = \{ \psi \in {\cal L}^2(dx), \psi(0) = \psi'(0) = 0 \}
\end{equation}
For $g \leq 3/4$,[Birmingham {\it et al.,},2001] the Hamiltonian
is not essentially self-adjoint, but it admits a one-parameter
family of self-adjoint extension labeled by a $U(1)$ parameter
$e^{iz}$, where $z$ is real, which labels the domains ${\cal D}_z$
of the extended Hamiltonian. The set ${\cal D}_z$ contains all the
vectors in ${\cal D}_0$, and vectors of the form $\psi_{+} +
e^{iz}\psi_{-}$.

For $g = - \frac{1}{4}$, and a given self-adjoint extension $z$,
there are infinite number of bound states:
\bea
\psi_n (x) &=& N_n \sqrt{x} K_0\left( \sqrt{E_n} x\right),\\
E_n &=& {\exp}\left[\frac{\pi}{2} (1 - 8n) {\cot}
\frac{z}{2}\right],
 \eea

 where $n$ is an integer, $N_n$ is a
normalization factor, and $K_{0}$ is the modified Bessel function.
The horizon, in this picture, is located at $x = 0$ where the wave
function $\psi _n$ vanishes. However, in a band-like region
$\Delta = [x_0(1-\delta ), x_0(1+\delta )]$, where $x_0 \sim
E_0^{-1/2}$ with $z > 0$,  all the eigenfunctions of $\hat{H}$
exibit a scaling behaviour of the type, $\psi _n \sim \sqrt{x}$ .

Formally, the three generators: $\hat{H}$ , the scaling operator
$\hat{D} = -(xp + px)/4$  and $\hat{K} = x^2/2$ obey the conformal
algebra commutation relations
\begin{equation}
  [\hat{H},\hat{D}] =i\hat{H},\;\;  [\hat{K},\hat{D}] = -i\hat{K},\;\;  [\hat{H},\hat{K}] =
  2i\hat{D},
\end{equation}

with a constant  quadratic Casimir operator
\begin{eqnarray}
C_2 & = & \frac{1}{2}(\hat{K}\hat{H} + \hat{H}\hat{K}) - \hat{D}^2
\nonumber \\& = & \frac{g}{4} - \frac{3}{16}.
\end{eqnarray}
 If $\hat{H}^{-1}$ exists [Klauder,1999], we can formally construct a time operator $\hat{t}$
 [Martinis and Mikuta,2003]
\begin{equation}
\hat{t} =  \frac{1}{2}( \hat{D}\hat{H}^{-1} +
\hat{H}^{-1}\hat{D}),
\end{equation}
which  obeys the required commutation relation, $[\hat{H},\hat{t}]
= i$. Although, both $\hat{H}$ and $\hat{D}$ separately can be
made self-adjoint operators it is not true for a
$\hat{t}$-operator which contains $\hat{H}^{-1}$ [Klauder,1999].
In the limit $g\longrightarrow 0$, we have $\hat{H}
\longrightarrow H_0 = p^2/2$ and $\hat{t}\longrightarrow t_0$
where
\begin{equation}
\hat{t}_0 = - \frac{1}{2}(xp^{-1} + p^{-1}x)
\end{equation}

is exactly the time-of-arrival operator of Aharonov and Bohm
[1961] The Aharonov-Bohn time operator $\hat{t}_0$ is not
self-adjoint and its eigenfunctions are not orthonormal.
\section{Conclusion}
In this paper, we have studied the  properties of a scalar field
in the near-horizon region of a massive Schwarzschild black hole.
The quantum Hamiltonian governing the near-horizon dynamics is
found to be  scale invariant and has the full conformal group as a
dynamical symmetry group. Using only the generators of the
conformal group, we constructed the time operator near-horizon.
The self-adjointness of $\hat{H}$ and $\hat{t}$ is also discussed.
\newpage

{\bf References}\\
\hspace{2cm}

 Aharonov, Y. \&  Bohm, D. [1961] "Time in
the Quantum Theory and the Uncertainty Relation for Time and
Energy," {\it Phys.
Rev.} {\bf 122}, 1649- .

Bai, H. \&  Yan, Mu-Lin [2004] {\it Quantum Horizon},
(gr-qc/0403064).

Birmingham, D. \& Gupta, K. S. \& Sen, S. [2001] "Near-Horiyon
Conformal Structure of Black Holes," {\it Phys. Lett.} {\bf B505},
191-196.
't Hooft,G. [2004] {\it Horizons} (gr-qc/0401027).

Klauder,J.R. [1999] {\it Noncanonical Quantization of Gravity I.
Foundation of Affine Quantum Gravity}, (gr-qc/9906013).

Martinis, M. \&  Mikuta, V. [2003] "Time Operator for a Singular
Quantum Oscillator" {\it Fizika} {\bf B12}, 285-290.

Nicolson, I. [1981] {\it Gravity, Black Holes,
and the Universe} (David and Charles, London).

Pauli, W. [1958] "Die allgemeinen Prinzipien der Wellenmechanik"
in {\it Encyclopedia of Physics}, ed. Fluegge, S.
(Springer-Verlag, Berlin)pp. 1-168.

\end{document}